\documentclass[11pt]{amsart}
\pdfoutput=1

\usepackage[utf8]{inputenc}

\usepackage[margin=3.25cm]{geometry}

\usepackage{amssymb, amsmath, amsthm, amsfonts}
\usepackage{graphicx}
\usepackage{color}
\usepackage{mathrsfs}
\usepackage{physics}
\usepackage{mathtools}
\usepackage{tikz}
\usepackage{faktor}
\usepackage{tikz-cd}
\usepackage{bm}
\usepackage[hidelinks]{hyperref}
\usepackage{csquotes}

\DeclareRobustCommand{\SkipTocEntry}[5]{}

\theoremstyle{remark}

\theoremstyle{remark}

\usepackage[UKenglish]{isodate}
\usepackage[UKenglish]{babel}

\renewcommand{\grad}{\nabla}

\title[Some minimal notes on notation and minima]{Some minimal notes on notation and minima: \\ A Comment on ``How Particular is the Physics of the Free Energy Principle?" by Aguilera, Millidge, Tschantz, and Buckley}

\author{Maxwell J D Ramstead}
\address{\parbox{\linewidth-12pt}{Wellcome Centre for Human Neuroimaging, University College London, London WC1N 3AR, UK}}
\address{VERSES Research Lab and Spatial Web Foundation, Los Angeles, CA, USA, 90016}
\email{maxwell.ramstead@verses.io}

\author{Dalton A R Sakthivadivel}
\address{\parbox{\linewidth-12pt}{Department of Mathematics, Department of Physics and Astronomy, Stony Brook University, Stony Brook, NY, USA, 11794-3651}}
\address{VERSES Research Lab and Spatial Web Foundation, Los Angeles, CA, USA, 90016}
\email{dalton.sakthivadivel@stonybrook.edu}
\urladdr{https://darsakthi.github.io}

\date{\today}

\let\oldtocsection=\tocsection

\let\oldtocsubsection=\tocsubsection

\renewcommand{\tocsection}[2]{\hspace{0em}\oldtocsection{#1}{#2}}
\renewcommand{\tocsubsection}[2]{\hspace{19.5pt}\oldtocsubsection{#1}{#2}}

\begin{document}

\maketitle

\begin{abstract}

We comment on a technical critique of the free energy principle in linear systems by Aguilera, Millidge, Tschantz, and Buckley, entitled ``How Particular is the Physics of the Free Energy Principle?" Aguilera and colleagues identify an ambiguity in the flow of the mode of a system, and we discuss the context for this ambiguity in earlier papers, and their proposal of a more adequate interpretation of these equations. Following that, we discuss a misinterpretation in their treatment of surprisal and variational free energy, especially with respect to their gradients and their minima. In sum, we argue that the results in the target paper are accurate and stand up to rigorous scrutiny; we also highlight that they, nonetheless, do not undermine the FEP. 

\end{abstract}

\date{\today}


\vskip0.5cm 

\section{Preliminary remarks}

We are delighted to comment on the interesting technical critique of the free energy principle (FEP) by Aguilera, Millidge, Tschantz, and Buckley. This critique is noteworthy for several reasons. Amongst them are the technical rigour and astuteness of the work done, for which \cite{aguilera2021} is distinct in the literature. We are especially pleased to see that the paper has provoked a dialogue which is both friendly and productive. Indeed, although we disagree with some aspects of the critique, we contend that, in the years since the paper was originally circulated as a preprint, our collective understanding of the core formalism and scope of the FEP has increased significantly---in no small part due to the conversations fostered by this critique. The community as a whole has undoubtedly benefited from this exchange, and we are grateful to have been part of it. 

In \cite{aguilera2021}, the authors present a detailed and rigorous analysis of the free energy principle. In this comment, we focus on two of the points made in \cite{aguilera2021}: one regarding marginal flows and their average, the other, regarding surprisal and variational free energy gradients, and their minima. We argue that, whilst the results in the target paper are accurate and stand up to rigorous scrutiny, they do not undermine the FEP. Lastly, we note that this was clearly by design: the work in the target paper was never meant to undermine the FEP. Rather, it was meant to test its applicability to\textemdash and informativeness about\textemdash linear systems.   

\section{Flows of averages and averages of flows}

Many of the core papers in the FEP literature (seem to) equate the average flow of a system to the flow of the average. It is certainly true that the average of the flow does not equal the flow of the average in general; nor even is this true generically. The authors are absolutely correct when they make this point. It is likely this error can be traced back to an unfortunate notational choice in \cite{parr}, where equation 3.5 therein reads
\[
\dot{\bm\eta}(b) = (Q - \Gamma) \grad \mathfrak{I}(\bm\eta, b). 
\]
This equation ought to have been written 
\begin{equation}\label{xxx}
\langle \dot\eta(b_t) \rangle = (Q - \Gamma) \grad \mathfrak{I}(\bm\eta_t, b_t),
\end{equation}
with $\langle \cdots \rangle$ denoting an expectation. The given decomposition of the flow is the deterministic component of a stochastic differential equation, arising from an averaging over the random fluctuations in the SDE perturbing its flow. That is, our equation \eqref{xxx} expresses (in clearer notation) the observation that the flow of the expected external state is predicated on the current blanket state and expected external state at a given time-point, and drifts based on the gradient of the surprisal of those variables, as calculated at that time point. 

That is precisely what is claimed in the FEP: that the flow of the mode is on average the drift component of an SDE for that flow. What is labelled as Assumption 3* in the target paper is not assumed by the FEP. Our equation \eqref{xxx}, which denotes equation 3.5 (ibid.) more properly, makes it apparent that one cannot read the flow operator as commuting with the expectation operator, which undermines Assumption 3* as a valid interpretation of the FEP. 

This should not, however, be construed as a mistake on the part of the authors. On the contrary, the authors reproduce the intended definition of marginal flow in \cite{parr}, which they label Assumption 3**. The point of critique raised here is only that their initial claim about the flow of the average and average of the flow in Assumption 3* is a \emph{non sequitur}. 

To summarise, the results of the target paper show that it would be nonsensical to interpret the FEP as asserting Assumption 3*, as it would lead to incorrect results. Luckily, the FEP does not depend on this assumption\textemdash despite, perhaps, being misleadingly written in some places. This is a purpose the critique serves decidedly well. It demonstrates rather clearly some pitfalls in the FEP literature, especially in the way it has been written in some core papers, and thus, the way it must not be read. 

Finally, since their analysis is predicated on Assumption 3** as though in retrospect, other results are independent of these statements.

\section{Gradients of surprisal and gradients of free energy}

The mishandling the definition of marginal flows in \cite{parr} touches indirectly on a second issue in the claims of the target paper, which we do take to be mistaken\textemdash namely, that the gradient of free energy ought to be informative about the average flow of external states, present in equation A9 of \cite{aguilera2021}.

It is obvious that these gradients exist in different spaces\textemdash hence the dual information geometry spoken of in \cite{afepfapp}\textemdash but to the authors’ credit, their argument is more subtle: the two gradients are linearly related when $\sigma$ is a linear function. However, here we note that linear transformations do not in general preserve the shape of a vector field, nor the shape of flows in that field. They simply map an identity element to an identity element. In particular, it is claimed in the target paper that when $\sigma$ acts linearly on the gradient of surprisal, it maps this vector field to the gradient of free energy linearly, and so we should be able to understand external states as flowing along the gradient of free energy just as well as along the gradient of surprisal. In fact, it is only the case that these vector fields share a minimum given by $\sigma$, and not that they reach those minima at the same time, nor for the same states. 

One can imagine a relatively simple counterexample to the foregoing statement that the gradient flow of one quantity is meaningful to a quantity which exhibits a linearly related gradient flow. Take, for instance, the planar vector fields $X = (-x, -y)$ and $Y = (y, -x)$, and the following linear transformation $T$ relating them:
\[
\begin{bmatrix}
\cos \frac{1}{2}\pi & -\sin \frac{1}{2}\pi \\[0.75em]
\sin \frac{1}{2}\pi & \cos \frac{1}{2}\pi
\end{bmatrix}
\begin{bmatrix}
-x \\[0.75em] -y
\end{bmatrix} = 
\begin{bmatrix}
y \\[0.75em] -x
\end{bmatrix}.
\]
We certainly have $X_{(0,0)} = (0,0)$ and $Y_{(0, 0)} = (0,0),$ and indeed, $T(0, 0) = (0,0).$ Hence, the zero point of $Y$ is the zero point of $Y$. However, the two gradients are very different, and their integral curves\textemdash the flows in that gradient\textemdash will also be quite different. This is not even to mention that when $T$ is non-linear\textemdash even an affine map, a linear map with a constant shift, will do\textemdash their minima will often occur at different states.  

In point of fact, exactly one such counterexample appears in the target paper: it is demonstrated that the gradient of $F(\bm\eta, b)$ does not resemble the gradient of $\mathfrak{I}(\bm\eta, b)$. Once again the authors are absolutely correct, but do not adequately connect their objection to the FEP. It is true, as they claim, that equating the two cannot be done; but nowhere does the FEP claim it can be done, nor does the FEP expect to draw insights from doing so. 

What the line of reasoning that the authors took \emph{would} allow us to say is that $\bm\eta$ flows on $\grad_\eta \mathfrak{I}(\bm\eta, b)$ whilst the \emph{linear transformation} of the flow of $\bm\eta,$ here $\grad \sigma^{-1}(\bm\eta),$ flows on $\grad_\mu \mathfrak{I}(\sigma(\bm\mu), b).$ This is equivalent to the existence of a relation 
\begin{equation}\label{flows}
(\grad_\mu\sigma)^{-1} \grad_\mu \mathfrak{I}(\sigma(\bm\mu), b) = \grad_\eta \mathfrak{I}(\bm\eta, b)
\end{equation}
between the two flows, which can indeed be produced by a simple application of the chain rule and $\sigma(\bm\mu) = \bm\eta$. Equation \eqref{flows} tells us that the flows in these vector fields are linearly related, but are emphatically \emph{not} identical. Respecting \eqref{flows}, the relation between flows holds in the same general sense as $T(x,y) = (x', y')$ for arbitrary pairs of vectors\textemdash not just zero points\textemdash but, it is concomitant on not equating one to the other at any point other than the zero point. Correspondingly, we note that $\bm\eta$ under the linear transformation $\sigma^{-1}$ is $\bm\mu,$ and that 
\[
\grad_\mu \mathfrak{I}(\sigma(\bm\mu), b) = \grad_\mu F(\bm\mu, b),
\]
recovering the equations in \cite{parr} \emph{and} illuminating the claim that
\[
\grad_\mu F(\bm\mu, b) \neq \grad_\eta \mathfrak{I}(\bm\eta, b)
\]
in \cite{aguilera2021}. The proper application of the synchronisation map $\sigma$ is known to generate the desired vector fields and flows, and in particular, relates the flow of $\bm\mu$ on the surprisal $\grad_\mu \mathfrak{I}(\sigma(\bm\mu), b)$ to a flow on the particular free energy $\grad_\mu F(\bm\mu, b),$ seen in \cite{dacosta}. 

As such, the FEP says nothing of free energy over external states, and avoids doing so by design. Here we arrive at an important critical remark: more generally, the FEP says nothing about whether systems actually do really minimise free energy gradients, or whether we can merely describe them as such; nor whether systems minimise their free energy, as opposed to merely tending towards a free energy minimum, and thus admitting an equivalent model of minimal free energy.

In closing, we are grateful to have been able to comment on this paper. We believe that the ensuing exchange will be part of a new, productive phase of development for the FEP. Consistent with the aim of the target paper, asking difficult technical questions of the FEP and searching for the answers is always a fruitful scientific exchange.

\bibliographystyle{alpha}
\bibliography{main}

\end{document}